\documentclass[reprint,prl,superscriptaddress]{revtex4-1}

\usepackage{subfigure}
\usepackage{graphicx}
\usepackage{adjustbox}
\usepackage{amssymb,amsmath}
\usepackage{upgreek}
\usepackage{float}

\usepackage{nicefrac}
\usepackage{url}

\begin{document}

\def\lsim{\mathrel{ \rlap{\raise.5ex\hbox{$<$}}
                      {\lower.5ex\hbox{$\sim$}}  } }
\def\gsim{\mathrel{ \rlap{\raise.5ex\hbox{$>$}}
                      {\lower.5ex\hbox{$\sim$}}  } }

\title{
Coherent Space-Charge to X-ray Up-Conversion in Gas }
\author{S. Li} \affiliation{SLAC National Accelerator Laboratory, Menlo park, California, 94025, USA}
\affiliation{Department of Physics, Stanford University, Stanford, California, 94305, USA}
\author{E. G. Champenois} \affiliation{Stanford PULSE Institute, SLAC National Accelerator Laboratory, Menlo park, California, 94025, USA}
\author{J. P. Cryan} \affiliation{Stanford PULSE Institute, SLAC National Accelerator Laboratory, Menlo park, California, 94025, USA}
\affiliation{Linac Coherent Light Source, SLAC National Accelerator Laboratory, Menlo park, California, 94025, USA}
\author{A. Marinelli} \affiliation{SLAC National Accelerator Laboratory, Menlo park, California, 94025, USA}

\begin{abstract}
We theoretically investigate the use of the intense transverse field created by a relativistic electron bunch to drive strong-field processes.
Such bunches can be easily implemented from beam shaping in existing electron accelerators. 
We focus on the process of strong-field driven high harmonic generation and calculate the single-atom dipole response to the space-charge field of relativistic tilted electron beam.
We compare the emitted radiation spectrum of the space-charge field to that of a few-cycle laser pulse and show that the space-charge field creates a spectrum which extends much further into the X-ray domain.
We apply simple classical trajectory analysis to understand this result. 
\end{abstract}

\maketitle

For a relativistic charged particle moving at constant velocity, the electric field along the direction of propagation is Lorentz invariant, whereas the transverse field is enhanced by a factor of $\gamma$, the Lorentz factor $1/\sqrt{1-(\nicefrac{v}{c})^2}$.
As a result, the electric field in the laboratory frame is flattened into an oblate or ``pancake'' distribution in the transverse plane. 
For the case of an electron beam, the effect turns into a collective space-charge field linearly proportional to the beam current with a peak amplitude $E_\perp \sim Z_0 I/2\pi r_b$, where $Z_0$ is the free-space impedance, $I$ is the beam current, and $r_b$ is the electron beam radius.
With modern electron accelerators driven by high-brightness photo-injectors it is possible to achieve transverse field amplitudes of hundreds of $\nicefrac{\mbox{GV}}{\mbox{m}}$ \cite{rosenzweig2011teravolt}, 
and beam durations on the order of a few femtoseconds~(fs). Strong space-charge fields have been employed to ionize plasmas in plasma-wakefield acceleration experiments \cite{corde2016high}, and the use of field-ionization from electron bunches has recently been proposed as a diagnostic tool for high-brightness electron bunches \cite{tarkeshian2018transverse}.

In this letter we theoretically investigate the use of the intense transverse field created by a relativistic electron bunch to drive strong field processes in atomic targets.
The interaction of an isolated atomic system with an intense electric field is one of the forefront problems in atomic, molecular, and optical physics.  
Over the past fifty years, the response of atomic, molecular, and solid-state systems to intense laser fields has received considerable attention. 
Initially the problem of tunnel ionization of free atoms subject to strong low-frequency fields was considered by Keldysh~\cite{keldysh1965ionization}.
With the development of high pulse energy, ultrafast laser systems, the availability of strong-laser fields, which approach (and exceed) the atomic unit of intensity has led to many new discoveries such as above-threshold ionization~(ATI)~\cite{agostini1979free}, above threshold dissociation~(ATD) in molecules~\cite{zavriyev1990ionization}, high harmonic generation~(HHG)~\cite{l1993high, mcpherson1987studies} and multiple ionization~\cite{walker1994precision}, while also posing a number of unanswered questions.  

For an isolated atomic or molecular system, the dynamics of an electron subject to an intense electromagnetic field are to good approximation described by a semi-classical theory that considers the classical motion of the electron in the field, neglecting the effect of the Coulomb potential~\cite{reiss1990complete,schafer1993above,corkum1993plasma,lewenstein1994theory}.
Using this strong field approximation (SFA) over recent decades, researchers have demonstrated valuable applications of controlling strongly driven electron dynamics.
Most notably, gas phase HHG has been used to generate attosecond pulse trains~\cite{paul2001observation} and isolated attosecond pulses~\cite{hentschel2001attosecond,krausz2009attosecond}, 
as well as a direct probe of strong-field driven attosecond electron motion with high harmonic spectroscopy~\cite{worner2010following}.
These previous works have only considered the case where the strong-field is derived from a laser source: we have yet to find a work that has taken advantage of the relativistic electron beam as the source for the strong field.
Here, we present the analysis of the single-atom dipole response to the transverse space-charge field of a relativistic electron beam as an investigation of the applicability of these beams to strong field physics.

In the semi-classical theory first described in~\cite{corkum1993plasma}, an intense laser field initially tunnel ionizes an electron from an atomic target. 
The tunnel ionized electron follows the classical trajectory for an electron subject to the strong laser field, with zero initial velocity.
For an oscillating electric field, electrons ionized at a time when the magnitude of the electric field is decreasing will eventually return to the ionic core, where it is possible to recombine and release the excess kinetic energy as a high energy photon. 
For a classical plane wave, the maximum kinetic energy of the returning electron is roughly $3.17U_p$, where $U_p=\nicefrac{E_0^2}{4\omega^2}$ is the ponderomotive or quiver energy~(in atomic units) of a free electron in an electric field with amplitude $E_0$.  
The three-step model predicts that the highest observable~(or cut-off) photon energy is given by $E_{\mbox{max}}=I_p+3.17U_p$, where $I_p$ is the ionization potential of the atom.  
The favorable $\lambda^2$-scaling of the ponderomotive potential can extend the HHG spectrum to soft X-ray wavelengths with the use of infrared drive lasers.


\begin{figure}
\centering
\includegraphics[width=0.5\textwidth]{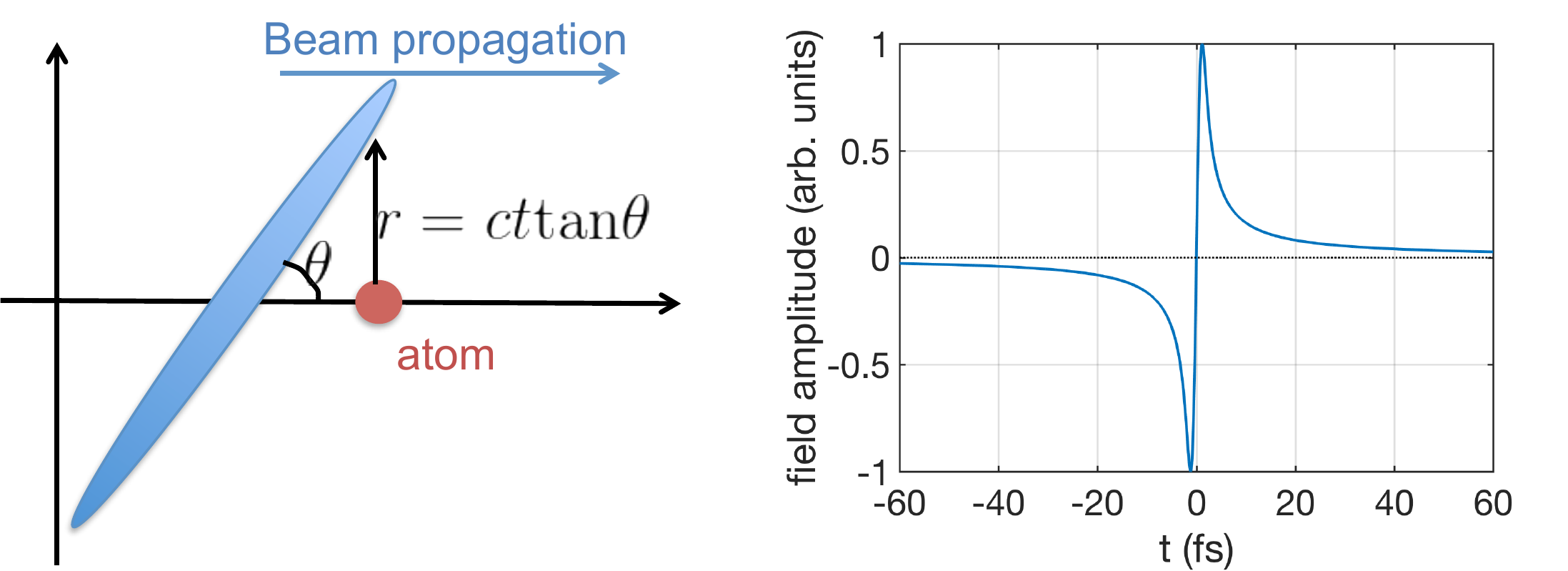}
\caption{Scheme for using the transverse space-charge field of a tilted electron beam for driving high harmonic generation. The left panel shows a tilted electron beam with an atom placed along the propagation direction of the beam. The right panel shows the electric field experienced by this atom, as a function of time.}
\label{fig:scheme}
\end{figure}

The space-charge field of an electron bunch as sampled by a static observer is typically a unipolar field, which is inadequate for the generation of harmonics in gas since it does not allow the electrons to return to the parent ion. 
Consider instead the space-charge field of a tilted relativistic electron beam measured at a point on-axis along the propagation direction, as depicted in Fig.~\ref{fig:scheme},
the space-charge field points radially outward at any point in space. 
The field experienced by an atom positioned along the symmetry axis of the electron bunch experiences a field that initially increases, and then quickly reverses sign at the pivot point of the electron bunch. 
We note that tilted electron bunches have been employed at the Linac Coherent Light Source~(LCLS) for the generation of two-color FEL pulses and can be generated with a passive dechirper \cite{lutman2016fresh}, with a transverse cavity, or using transverse dispersion with a chirped electron bunch~\cite{zholents2005method,hemsing2014beam,prat2015efficient,huang2017generating}.
The electric field strength of the tilted bunch can easily reach up to tens of ${\mbox{GV}}/{\mbox{m}}$. 
In what follows, we consider the space-charge field of a tilted electron bunch as a driving field for HHG.
We find an interesting result, that the characteristic temporal structure of such a driving field creates a broader HHG spectrum than is expected from a similar few-cycle laser pulse.

Assuming a Gaussian transverse distribution of the electrons, the space-charge field can be written as,
\begin{equation}
E(r)=\left.\left. \left(\frac{Z_0}{2\pi}\right) \frac{I_b}{r} \right( 1-e^{-\frac{r^2}{2\sigma_r^2}} \right),
\label{eq:E_r}
\end{equation}
where $Z_0$ is the free-space impedance, $I_b$ is the beam current, $\sigma_r$ is the root-mean-square width of the slice, and the $r$ coordinate is defined as the distance from the axis of beam propagation to the center of the beam slice under consideration. When the electron beam is tilted by an angle of $\theta$, we can write $r=ct\mbox{tan}\theta$, where we assume that the electron beam is traveling at highly relativistic speeds close to the speed of light, $c$. 
We can rewrite Eq.~\ref{eq:E_r} as 
\begin{equation}
E(t)=\frac{E_0}{t/\sigma_t}\left(1-\mbox{exp}(-\frac{t^2}{2\sigma_t^2})\right)F(t)
\label{eq:E_r_simplified}
\end{equation}
where we have defined $\sigma_t=\nicefrac{\sigma_r}{c\rm{tan}\theta}$ and grouped the prefactors into an amplitude $E_0$.
The function $F(t)$ in Eq.~\ref{eq:E_r_simplified} is a characteristic current function of the electron bunch. This expression is valid if the electron bunch length in the rest frame is much larger than the transverse beam size $L_b \gamma >> \sigma_r$, and if the tilt angle in the rest frame is much smaller than unity.
Thus we transform the expression of the electric field with variables $\theta$ and $\sigma_r$ into an expression with two independent variables $E_0$ and $\sigma_t$. The variable $\sigma_t$ is linearly proportional to the wavelength $\lambda$ of the field, which we define as the central frequency of the Fourier transformed field (Eq.~\ref{eq:E_r_simplified}).
For the example field profile in Fig.~\ref{fig:scheme}, $\lambda=\rm{2}~{\mu m}$ (or $\sigma_t=0.11\lambda/c$ with the proportionality factor found empirically) and the field amplitude is taken to be unity.
%

\begin{figure}
\centering
\includegraphics[width=0.5\textwidth]{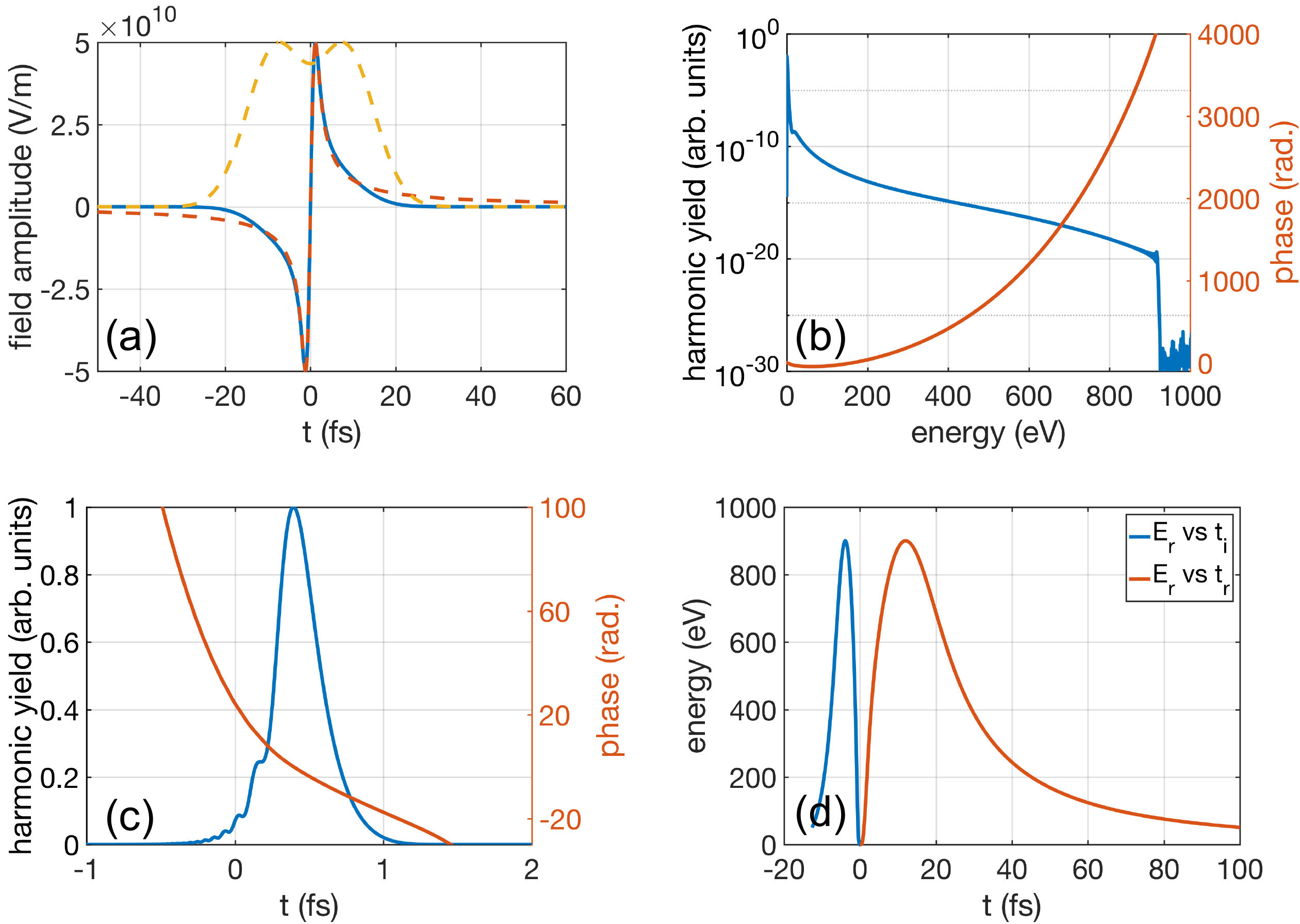}
\caption{Space-charge field of a tilted electron beam~(a) with a beam filter constructed by a double Gaussian function (yellow curve). The dashed red curve is the field calculated by Eqs.~\ref{eq:E_r} and \ref{eq:E_r_simplified}. The composite field is shown as a solid blue curve. Panel~(b) shows the Fourier transform amplitude~(blue) and phase~(red) of the time-dependent dipole moment from quantum mechanical calculation, using the electric field shown in~(a). Panel~(c) shows the amplitude~(blue) and phase~(red) of the high-pass filtered~(see text) dipole moment in the time-domain. Panel~(d) shows the results of trajectory simulations for the  electron return energy $E_r$ as a function of ionization time $t_i$ and return time $t_r$.}
\label{fig:tiltbeam}
\end{figure}

The general shape of the electric field from the tilted electron beam resembles that of a sinusoidal wave with zero crossing on axis, except the former has a non-zero tail due to the $\frac{1}{t}$ factor in front. 
Realistically, we implement a beam filter which takes account of the finite longitudinal size of the electron beam that effectively kills the long tails. 
It is common to have a double horn structure of the electron beam coming from an electron accelerator~\cite{behrens14etal}, so we construct the beam filter by superposing two Gaussian functions together with a time separation equal to the bunch length. 
By multiplying the beam filter with the electric field in Eq.~\ref{eq:E_r_simplified} we get a more realistic representation of the field generated from a tilted electron beam.
Note that the effect of the beam filter is mostly to provide a finite bunch length constraint. The result in this letter is not sensitive to the specific shape of the beam filter. We choose the double Gaussin filter due to its similarity to the double horn profile commonly seen at LCLS.

In order to approximate the HHG spectrum we calculate the time-dependent dipole moment of a single atom placed along the axis of beam propagation, as depicted in Fig.~\ref{fig:scheme}.
Our quantum mechanical calculations are made in the SFA using an open source software package~\cite{HHGmaxcode}.
The field shown in Fig.~\ref{fig:tiltbeam} is the transverse field of an electron beam tilted by an angle $\theta=85.8^{\circ}$ with $\sigma_r=3~\mu$m.
For a peak current of a few kilo-amps, the peak field amplitude can reach 50~GV/m.
Taking the Fourier transform of the temporal field we find the wavelength to be 2~$\mu$m. 
The double Gaussian beam filter has a root-mean-square width of $2~\mu$m and a peak separation of 5~$\mu$m.
The calculated single-atom HHG spectrum exhibits a nearly structureless plateau region out to more than 900~eV.
To produce a time-domain dipole response, the low frequency ($<20$~eV) contributions are first filtered out by a half-Gaussian, high-pass filter, centered at 20~eV and then Fourier transformed.
The modulus squared of the time-domain dipole response (Fig.~\ref{fig:tiltbeam}~(c)) shows that these X-rays are emitted in a single burst.
The emitted X-ray pulse has a duration of 316~as~(FWHM), which is substantially longer than the Fourier transform limit.
This is due to the large group delay dispersion (or chirp) which results from the HHG process~\cite{doumy2009attosecond}.

\begin{figure}
\centering
\includegraphics[width=0.5\textwidth]{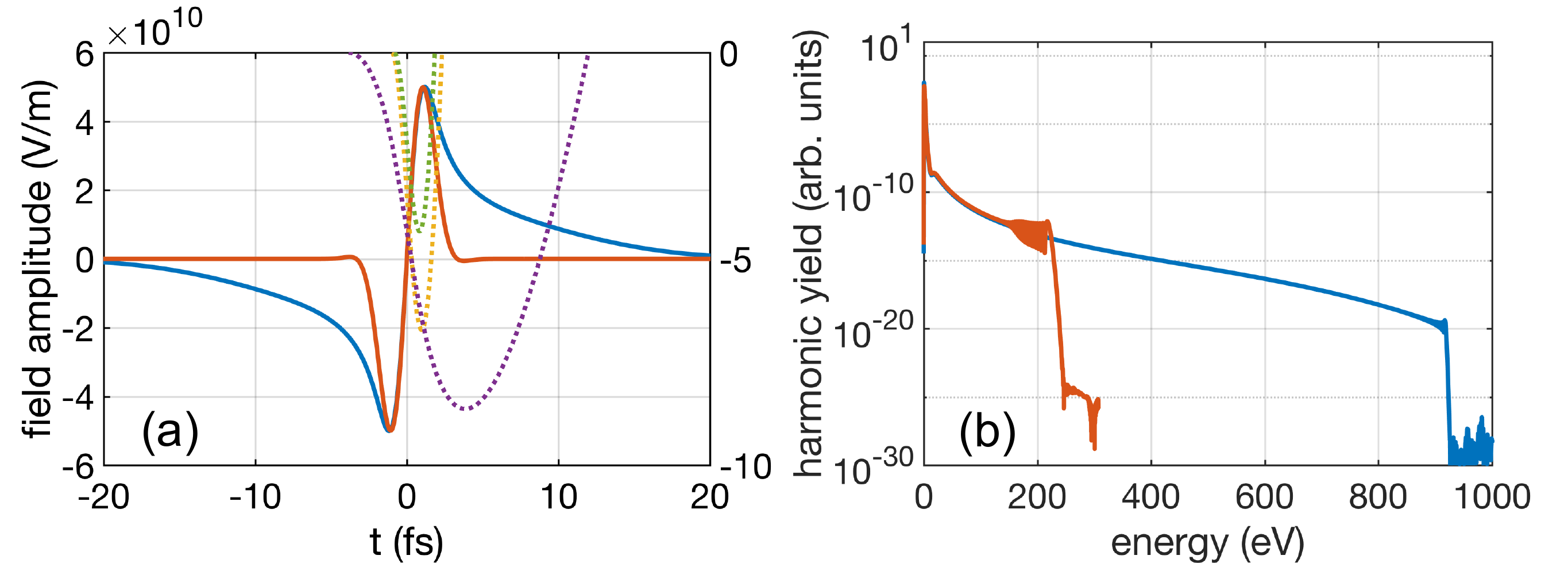}
\caption{Comparison of the emitted radiation spectrum for both the space-charge field~(blue) and a  single-cycle laser pulse~(red). Panel~(a) shows the time domain electric field for the two pulses. Panel~(b) shows the amplitude of the Fourier transform of the simulated time-dependent dipole moment. 
Panel~(a) also shows the results of trajectory analysis for the trajectory which corresponds to an electron with maximum return energy driven by the single-cycle sinusoidal wave~(dashed, yellow), and the space-charge field~(dashed, purple).  
The amplitude of dashed purple curve has been scaled down by a factor of 10 to better fit in the figure.
We also include the trajectory~(dashed, green) that corresponds to an electron driven by the space-charge field that returns with the maximum energy for the single-cycle pulse. }
\label{fig:compare2sine}
\end{figure}

As a comparison with HHG from more conventional ultrafast laser pulses, Fig.~\ref{fig:compare2sine} shows the dipole spectrum calculated for the field considered in Fig.~\ref{fig:tiltbeam} along with a single-cycle sinusoidal laser pulse with a central wavelength of $2~\mu$m, identical to the central wavelength of the space-charge field.
The two spectra look qualitatively similar below 200~eV, however the spectrum of the dipole moment from the space-charge field extends to much higher photon energies.
To better understand the origin of the differences in Fig.~\ref{fig:compare2sine} and for overall understanding of the HHG process for space-charge beams, we turn to a classical trajectory model described in the introduction. 
In the semi-classical three-step model, the energy of the emitted photon,$E_{\gamma}$, is given by the sum of the energy of the returning electron, $E_R$ and the ionization potential of the target, $I_p$~\cite{corkum1993plasma},
\begin{equation}
E_{\gamma} = E_R + I_p.
\end{equation}
In this model, the energy of the returning electron is given by,
\begin{equation}
E_R(t_i) = \frac{p(t_r;t_i)^2}{2} [\mbox{in a.u.}],
\end{equation}
where $\vec{p}(t;t_i)= \alpha (\vec{A}(t_i) - \vec{A}(t))$, is the momentum of an electron~(in a.u.) ionized at time $t_i$. 
$\vec{A}(t)$ is the vector potential of the incident electric-field, $\vec{E}(t) = -\alpha\partial_t \vec{A}(t)$
. 
The ionized electron will return to the ionic core at time $t_r$ satisfying the relation, 
\begin{equation}
\int_{t_i}^{t_r(t_i)} dt' p(t';t_i) = 0,
\end{equation}
Figure~\ref{fig:tiltbeam}~(d) shows the result of the three-step model analysis.
The maximum return energy for the ionized electron is $\sim900$~eV. 
When added to the ionization potential of hydrogen~(13.6~eV), this gives a cut-off energy of $\sim914$~eV, in agreement with the quantum mechanical simulation~(Fig.~\ref{fig:tiltbeam}). 
This is because the main contribution to the dipole moment comes from the quantum paths whose quasiclassical action is stationary, and therefore represents a classical trajectory~\cite{lewenstein1994theory}. 

Figure~\ref{fig:compare2sine}~(a) shows three examples of classical trajectories for the ionized electron in the external electric field.
The dashed yellow curve shows the electron trajectory corresponding to the maximum return energy for the single-cycle laser field~(red), whereas the dashed purple curve shows the trajectory corresponding to the maximum return energy for an electron subject to the space-charge field (blue).
The dashed green curve shows the trajectory for an ionized electron subject to the space-charge field that has the same return energy as the single-cycle field cut-off energy.
The space-charge field driven trajectories corresponding to the highest recombination energies make use of the temporal tails of the field.
The leading edge leads to trajectories with large electronic excursions, while the trailing edge further increases the recombination energy of these long trajectories in comparison to laser-driven trajectories where the field goes to zero (single cycle) or even decelerates the electron (multi-cycle). 

\begin{figure}
\centering
\includegraphics[width=0.5\textwidth]{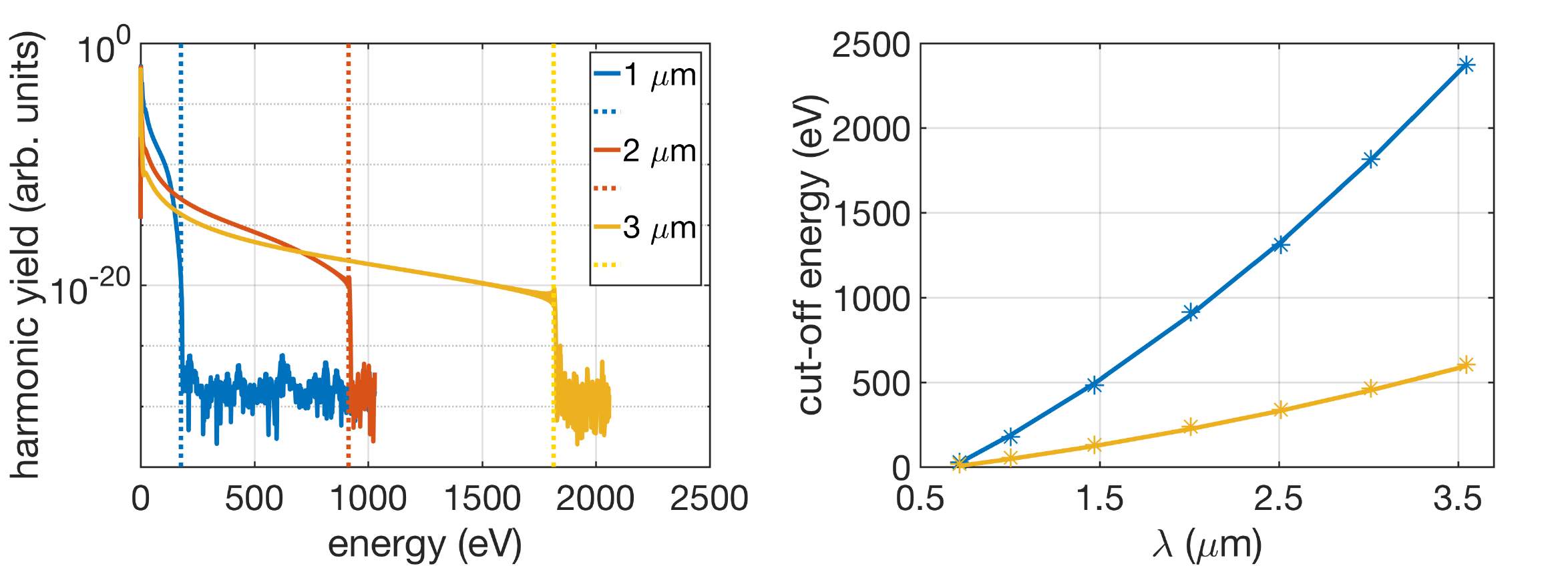}
\caption{Simulated radiation spectrum for different wavelengths of the tilted beam field~(see text for a description of how the central wavelength is varied).
The left plot shows the amplitude of the Fourier transform of the dipole response for a space-charge field with central wavelengths of 1~$\mu$m~(blue), 2~$\mu$m~(red), and 3~$\mu$m~(yellow), and $E_0=50~\mbox{GV/m}$. 
The vertical dotted lines indicate the cut-off frequency calculated from the classical three-step model. 
The right panel shows the cut-off energy as a function of driving field wavelength for $E_0=50~\mbox{GV/m}$~(blue stars) and $E_0=25~\mbox{GV/m}$~(yellow stars). 
The line shows the result of a second-order polynomial fit: 
$E_{\mbox{max}}[\mbox{eV}]=(E_0[\mbox{GV/m}])^2\left(0.045\lambda[\mu \mbox{m}]^2+0.160\lambda[\mu\mbox{m}]-0.126\right).$ 
}
\label{fig:scan_lambda}
\end{figure}

We have also investigated the effect of the space-charge field wavelength on the single-atom dipole response, by varying the length of the electron bunch, $\sigma_t$, while keeping the beam filter unchanged.
Figure~\ref{fig:scan_lambda} shows the case of 1~$\mu$m, 2$~\mu$m, and 3~$\mu$m central wavelength. 
Increasing the electron bunch length (or wavelength) results in an extension of the high energy cut-off, similar to conventional laser driven HHG.
For multi-cycle laser pulses the cut-off energy scales quadratically with the drive laser wavelength.
However, the space-charge field shows a wavelength scaling that is dominantly linear.
The linear dependence is a consequence of the fact that we only change $\sigma_t$, and do not scale the beam filter function, which corresponds to varying the tilt angle and transverse beam size while leaving the current profile unchanged. 
Conversely, scaling the bunch duration together with $\sigma_t$ results in the cutoff energy scaling as  $E_{max} \propto \lambda^2$, consistently with the behavior of laser-based HHG. 
Additionally, we find that the cut-off energy scales linearly with the field intensity $|E_0|^2$ in both cases.

In conclusion, we have proposed to use the extreme fields generated by the space-charge field of a tilted electron bunch to drive strong-field processes in atomic systems.
As a first example, we consider strong-field driven high harmonic generation, which is a hallmark problem in strong field atomic physics. 
We implemented quantum mechanical simulations of the HHG process in the strong-field approximation with realistic beam parameters and compared these results to more conventional few-cycle laser pulse HHG. 
We find that the unique temporal shape of the space-charge field has interesting consequences on the free electron wavepacket, which leads to extension in the photon energy cutoff. 


Although we have focused on the single-atom dipole response, propagation effects will also play a crucial role in the spectral shaping of the generated light and the overall efficiency of the process.
Phase matching considerations for a relativistic electron beam are quite different than those for an infrared driving pulse.
Preliminary calculations assuming a constant temporal profile of the space-charge field indicate that coherent buildup over $>$10~bar$\cdot$mm can be reached in the soft X-ray region, limited by the superluminal phase velocity of the emitted X-rays.
Interestingly, the increase of the index of refraction at atomic core resonances could help phase match the HHG process locally.
However, the effect of the gas target and of the induced plasma on the electron bunch may also need to be considered.

A potentially useful consequence of HHG driven directly by a relativistic beam is the possibility of creating a broad-bandwidth, sub-femtosecond X-ray pulse that are co-timed to the electrons with attosecond precision.
These HHG pulses could be propagated along with the relativistic beam through an undulator line to create tunable narrow-band X-ray pulses which are concomitant with an X-ray continuum.  

In addition to being a useful process for creating an X-ray continuum, we believe that the transverse field of an electron bunch is a useful tool for many strong-field physics problems. 
For example one could use space-charge fields to drive above threshold ionization, above threshold dissociation in molecular systems and multiple ionization. 

The authors would like to acknowledge P. Bucksbaum, D. Reis, S. Ghimire, J. B. Rosenzweig, Z. Huang, R. Ischebeck, R. Tarkeshian, and P. Krejcik for useful discussions and suggestions. This work was supported by the U.S. Department of Energy, Office of Science, under Contract No. DE-AC02-76SF00515 and FWP 100317. SL acknowledges support from the Robert H. Siemann Graduate Fellowships in Physics. The work by JPC was supported by the U.S. Department of Energy (DOE), Office of Science, Basic Energy Sciences (BES), Chemical Sciences, Geosciences, and Biosciences Division.

\bibliographystyle{apsrev4-1} 
\bibliography{all_ref}{}  

\begin{thebibliography}{25}%
\makeatletter
\providecommand \@ifxundefined [1]{%
 \@ifx{#1\undefined}
}%
\providecommand \@ifnum [1]{%
 \ifnum #1\expandafter \@firstoftwo
 \else \expandafter \@secondoftwo
 \fi
}%
\providecommand \@ifx [1]{%
 \ifx #1\expandafter \@firstoftwo
 \else \expandafter \@secondoftwo
 \fi
}%
\providecommand \natexlab [1]{#1}%
\providecommand \enquote  [1]{``#1''}%
\providecommand \bibnamefont  [1]{#1}%
\providecommand \bibfnamefont [1]{#1}%
\providecommand \citenamefont [1]{#1}%
\providecommand \href@noop [0]{\@secondoftwo}%
\providecommand \href [0]{\begingroup \@sanitize@url \@href}%
\providecommand \@href[1]{\@@startlink{#1}\@@href}%
\providecommand \@@href[1]{\endgroup#1\@@endlink}%
\providecommand \@sanitize@url [0]{\catcode `\\12\catcode `\$12\catcode
  `\&12\catcode `\#12\catcode `\^12\catcode `\_12\catcode `\%12\relax}%
\providecommand \@@startlink[1]{}%
\providecommand \@@endlink[0]{}%
\providecommand \url  [0]{\begingroup\@sanitize@url \@url }%
\providecommand \@url [1]{\endgroup\@href {#1}{\urlprefix }}%
\providecommand \urlprefix  [0]{URL }%
\providecommand \Eprint [0]{\href }%
\providecommand \doibase [0]{http://dx.doi.org/}%
\providecommand \selectlanguage [0]{\@gobble}%
\providecommand \bibinfo  [0]{\@secondoftwo}%
\providecommand \bibfield  [0]{\@secondoftwo}%
\providecommand \translation [1]{[#1]}%
\providecommand \BibitemOpen [0]{}%
\providecommand \bibitemStop [0]{}%
\providecommand \bibitemNoStop [0]{.\EOS\space}%
\providecommand \EOS [0]{\spacefactor3000\relax}%
\providecommand \BibitemShut  [1]{\csname bibitem#1\endcsname}%
\let\auto@bib@innerbib\@empty
\bibitem [{\citenamefont {Rosenzweig}\ \emph {et~al.}(2011)\citenamefont
  {Rosenzweig}, \citenamefont {Andonian}, \citenamefont {Bucksbaum},
  \citenamefont {Ferrario}, \citenamefont {Full}, \citenamefont {Fukusawa},
  \citenamefont {Hemsing}, \citenamefont {Hidding}, \citenamefont {Hogan},
  \citenamefont {Krejcik} \emph {et~al.}}]{rosenzweig2011teravolt}%
  \BibitemOpen
  \bibfield  {author} {\bibinfo {author} {\bibfnamefont {J.}~\bibnamefont
  {Rosenzweig}}, \bibinfo {author} {\bibfnamefont {G.}~\bibnamefont
  {Andonian}}, \bibinfo {author} {\bibfnamefont {P.}~\bibnamefont {Bucksbaum}},
  \bibinfo {author} {\bibfnamefont {M.}~\bibnamefont {Ferrario}}, \bibinfo
  {author} {\bibfnamefont {S.}~\bibnamefont {Full}}, \bibinfo {author}
  {\bibfnamefont {A.}~\bibnamefont {Fukusawa}}, \bibinfo {author}
  {\bibfnamefont {E.}~\bibnamefont {Hemsing}}, \bibinfo {author} {\bibfnamefont
  {B.}~\bibnamefont {Hidding}}, \bibinfo {author} {\bibfnamefont
  {M.}~\bibnamefont {Hogan}}, \bibinfo {author} {\bibfnamefont
  {P.}~\bibnamefont {Krejcik}},  \emph {et~al.},\ }\href@noop {} {\bibfield
  {journal} {\bibinfo  {journal} {Nuclear Instruments and Methods in Physics
  Research Section A: Accelerators, Spectrometers, Detectors and Associated
  Equipment}\ }\textbf {\bibinfo {volume} {653}},\ \bibinfo {pages} {98}
  (\bibinfo {year} {2011})}\BibitemShut {NoStop}%
\bibitem [{\citenamefont {Corde}\ \emph {et~al.}(2016)\citenamefont {Corde},
  \citenamefont {Adli}, \citenamefont {Allen}, \citenamefont {An},
  \citenamefont {Clarke}, \citenamefont {Clausse}, \citenamefont {Clayton},
  \citenamefont {Delahaye}, \citenamefont {Frederico}, \citenamefont {Gessner}
  \emph {et~al.}}]{corde2016high}%
  \BibitemOpen
  \bibfield  {author} {\bibinfo {author} {\bibfnamefont {S.}~\bibnamefont
  {Corde}}, \bibinfo {author} {\bibfnamefont {E.}~\bibnamefont {Adli}},
  \bibinfo {author} {\bibfnamefont {J.}~\bibnamefont {Allen}}, \bibinfo
  {author} {\bibfnamefont {W.}~\bibnamefont {An}}, \bibinfo {author}
  {\bibfnamefont {C.}~\bibnamefont {Clarke}}, \bibinfo {author} {\bibfnamefont
  {B.}~\bibnamefont {Clausse}}, \bibinfo {author} {\bibfnamefont
  {C.}~\bibnamefont {Clayton}}, \bibinfo {author} {\bibfnamefont
  {J.}~\bibnamefont {Delahaye}}, \bibinfo {author} {\bibfnamefont
  {J.}~\bibnamefont {Frederico}}, \bibinfo {author} {\bibfnamefont
  {S.}~\bibnamefont {Gessner}},  \emph {et~al.},\ }\href@noop {} {\bibfield
  {journal} {\bibinfo  {journal} {Nature communications}\ }\textbf {\bibinfo
  {volume} {7}},\ \bibinfo {pages} {11898} (\bibinfo {year}
  {2016})}\BibitemShut {NoStop}%
\bibitem [{\citenamefont {Tarkeshian}\ \emph {et~al.}(2018)\citenamefont
  {Tarkeshian}, \citenamefont {Vay}, \citenamefont {Lehe}, \citenamefont
  {Schroeder}, \citenamefont {Esarey}, \citenamefont {Feurer},\ and\
  \citenamefont {Leemans}}]{tarkeshian2018transverse}%
  \BibitemOpen
  \bibfield  {author} {\bibinfo {author} {\bibfnamefont {R.}~\bibnamefont
  {Tarkeshian}}, \bibinfo {author} {\bibfnamefont {J.}~\bibnamefont {Vay}},
  \bibinfo {author} {\bibfnamefont {R.}~\bibnamefont {Lehe}}, \bibinfo {author}
  {\bibfnamefont {C.}~\bibnamefont {Schroeder}}, \bibinfo {author}
  {\bibfnamefont {E.}~\bibnamefont {Esarey}}, \bibinfo {author} {\bibfnamefont
  {T.}~\bibnamefont {Feurer}}, \ and\ \bibinfo {author} {\bibfnamefont
  {W.}~\bibnamefont {Leemans}},\ }\href@noop {} {\bibfield  {journal} {\bibinfo
   {journal} {Physical Review X}\ }\textbf {\bibinfo {volume} {8}},\ \bibinfo
  {pages} {021039} (\bibinfo {year} {2018})}\BibitemShut {NoStop}%
\bibitem [{\citenamefont {Keldysh}\ \emph {et~al.}(1965)\citenamefont {Keldysh}
  \emph {et~al.}}]{keldysh1965ionization}%
  \BibitemOpen
  \bibfield  {author} {\bibinfo {author} {\bibfnamefont {L.}~\bibnamefont
  {Keldysh}} \emph {et~al.},\ }\href@noop {} {\bibfield  {journal} {\bibinfo
  {journal} {Sov. Phys. JETP}\ }\textbf {\bibinfo {volume} {20}},\ \bibinfo
  {pages} {1307} (\bibinfo {year} {1965})}\BibitemShut {NoStop}%
\bibitem [{\citenamefont {Agostini}\ \emph {et~al.}(1979)\citenamefont
  {Agostini}, \citenamefont {Fabre}, \citenamefont {Mainfray}, \citenamefont
  {Petite},\ and\ \citenamefont {Rahman}}]{agostini1979free}%
  \BibitemOpen
  \bibfield  {author} {\bibinfo {author} {\bibfnamefont {P.}~\bibnamefont
  {Agostini}}, \bibinfo {author} {\bibfnamefont {F.}~\bibnamefont {Fabre}},
  \bibinfo {author} {\bibfnamefont {G.}~\bibnamefont {Mainfray}}, \bibinfo
  {author} {\bibfnamefont {G.}~\bibnamefont {Petite}}, \ and\ \bibinfo {author}
  {\bibfnamefont {N.~K.}\ \bibnamefont {Rahman}},\ }\href@noop {} {\bibfield
  {journal} {\bibinfo  {journal} {Physical Review Letters}\ }\textbf {\bibinfo
  {volume} {42}},\ \bibinfo {pages} {1127} (\bibinfo {year}
  {1979})}\BibitemShut {NoStop}%
\bibitem [{\citenamefont {Zavriyev}\ \emph {et~al.}(1990)\citenamefont
  {Zavriyev}, \citenamefont {Bucksbaum}, \citenamefont {Muller},\ and\
  \citenamefont {Schumacher}}]{zavriyev1990ionization}%
  \BibitemOpen
  \bibfield  {author} {\bibinfo {author} {\bibfnamefont {A.}~\bibnamefont
  {Zavriyev}}, \bibinfo {author} {\bibfnamefont {P.~H.}\ \bibnamefont
  {Bucksbaum}}, \bibinfo {author} {\bibfnamefont {H.~G.}\ \bibnamefont
  {Muller}}, \ and\ \bibinfo {author} {\bibfnamefont {D.~W.}\ \bibnamefont
  {Schumacher}},\ }\href@noop {} {\bibfield  {journal} {\bibinfo  {journal}
  {Physical Review A}\ }\textbf {\bibinfo {volume} {42}},\ \bibinfo {pages}
  {5500} (\bibinfo {year} {1990})}\BibitemShut {NoStop}%
\bibitem [{\citenamefont {L’Huillier}\ and\ \citenamefont
  {Balcou}(1993)}]{l1993high}%
  \BibitemOpen
  \bibfield  {author} {\bibinfo {author} {\bibfnamefont {A.}~\bibnamefont
  {L’Huillier}}\ and\ \bibinfo {author} {\bibfnamefont {P.}~\bibnamefont
  {Balcou}},\ }\href@noop {} {\bibfield  {journal} {\bibinfo  {journal}
  {Physical Review Letters}\ }\textbf {\bibinfo {volume} {70}},\ \bibinfo
  {pages} {774} (\bibinfo {year} {1993})}\BibitemShut {NoStop}%
\bibitem [{\citenamefont {McPherson}\ \emph {et~al.}(1987)\citenamefont
  {McPherson}, \citenamefont {Gibson}, \citenamefont {Jara}, \citenamefont
  {Johann}, \citenamefont {Luk}, \citenamefont {McIntyre}, \citenamefont
  {Boyer},\ and\ \citenamefont {Rhodes}}]{mcpherson1987studies}%
  \BibitemOpen
  \bibfield  {author} {\bibinfo {author} {\bibfnamefont {A.}~\bibnamefont
  {McPherson}}, \bibinfo {author} {\bibfnamefont {G.}~\bibnamefont {Gibson}},
  \bibinfo {author} {\bibfnamefont {H.}~\bibnamefont {Jara}}, \bibinfo {author}
  {\bibfnamefont {U.}~\bibnamefont {Johann}}, \bibinfo {author} {\bibfnamefont
  {T.~S.}\ \bibnamefont {Luk}}, \bibinfo {author} {\bibfnamefont
  {I.}~\bibnamefont {McIntyre}}, \bibinfo {author} {\bibfnamefont
  {K.}~\bibnamefont {Boyer}}, \ and\ \bibinfo {author} {\bibfnamefont {C.~K.}\
  \bibnamefont {Rhodes}},\ }\href@noop {} {\bibfield  {journal} {\bibinfo
  {journal} {JOSA B}\ }\textbf {\bibinfo {volume} {4}},\ \bibinfo {pages} {595}
  (\bibinfo {year} {1987})}\BibitemShut {NoStop}%
\bibitem [{\citenamefont {Walker}\ \emph {et~al.}(1994)\citenamefont {Walker},
  \citenamefont {Sheehy}, \citenamefont {DiMauro}, \citenamefont {Agostini},
  \citenamefont {Schafer},\ and\ \citenamefont
  {Kulander}}]{walker1994precision}%
  \BibitemOpen
  \bibfield  {author} {\bibinfo {author} {\bibfnamefont {B.}~\bibnamefont
  {Walker}}, \bibinfo {author} {\bibfnamefont {B.}~\bibnamefont {Sheehy}},
  \bibinfo {author} {\bibfnamefont {L.~F.}\ \bibnamefont {DiMauro}}, \bibinfo
  {author} {\bibfnamefont {P.}~\bibnamefont {Agostini}}, \bibinfo {author}
  {\bibfnamefont {K.~J.}\ \bibnamefont {Schafer}}, \ and\ \bibinfo {author}
  {\bibfnamefont {K.~C.}\ \bibnamefont {Kulander}},\ }\href@noop {} {\bibfield
  {journal} {\bibinfo  {journal} {Physical review letters}\ }\textbf {\bibinfo
  {volume} {73}},\ \bibinfo {pages} {1227} (\bibinfo {year}
  {1994})}\BibitemShut {NoStop}%
\bibitem [{\citenamefont {Reiss}(1990)}]{reiss1990complete}%
  \BibitemOpen
  \bibfield  {author} {\bibinfo {author} {\bibfnamefont {H.}~\bibnamefont
  {Reiss}},\ }\href@noop {} {\bibfield  {journal} {\bibinfo  {journal}
  {Physical Review A}\ }\textbf {\bibinfo {volume} {42}},\ \bibinfo {pages}
  {1476} (\bibinfo {year} {1990})}\BibitemShut {NoStop}%
\bibitem [{\citenamefont {Schafer}\ \emph {et~al.}(1993)\citenamefont
  {Schafer}, \citenamefont {Yang}, \citenamefont {DiMauro},\ and\ \citenamefont
  {Kulander}}]{schafer1993above}%
  \BibitemOpen
  \bibfield  {author} {\bibinfo {author} {\bibfnamefont {K.}~\bibnamefont
  {Schafer}}, \bibinfo {author} {\bibfnamefont {B.}~\bibnamefont {Yang}},
  \bibinfo {author} {\bibfnamefont {L.}~\bibnamefont {DiMauro}}, \ and\
  \bibinfo {author} {\bibfnamefont {K.}~\bibnamefont {Kulander}},\ }\href@noop
  {} {\bibfield  {journal} {\bibinfo  {journal} {Physical review letters}\
  }\textbf {\bibinfo {volume} {70}},\ \bibinfo {pages} {1599} (\bibinfo {year}
  {1993})}\BibitemShut {NoStop}%
\bibitem [{\citenamefont {Corkum}(1993)}]{corkum1993plasma}%
  \BibitemOpen
  \bibfield  {author} {\bibinfo {author} {\bibfnamefont {P.~B.}\ \bibnamefont
  {Corkum}},\ }\href@noop {} {\bibfield  {journal} {\bibinfo  {journal}
  {Physical Review Letters}\ }\textbf {\bibinfo {volume} {71}},\ \bibinfo
  {pages} {1994} (\bibinfo {year} {1993})}\BibitemShut {NoStop}%
\bibitem [{\citenamefont {Lewenstein}\ \emph {et~al.}(1994)\citenamefont
  {Lewenstein}, \citenamefont {Balcou}, \citenamefont {Ivanov}, \citenamefont
  {L’huillier},\ and\ \citenamefont {Corkum}}]{lewenstein1994theory}%
  \BibitemOpen
  \bibfield  {author} {\bibinfo {author} {\bibfnamefont {M.}~\bibnamefont
  {Lewenstein}}, \bibinfo {author} {\bibfnamefont {P.}~\bibnamefont {Balcou}},
  \bibinfo {author} {\bibfnamefont {M.~Y.}\ \bibnamefont {Ivanov}}, \bibinfo
  {author} {\bibfnamefont {A.}~\bibnamefont {L’huillier}}, \ and\ \bibinfo
  {author} {\bibfnamefont {P.~B.}\ \bibnamefont {Corkum}},\ }\href@noop {}
  {\bibfield  {journal} {\bibinfo  {journal} {Physical Review A}\ }\textbf
  {\bibinfo {volume} {49}},\ \bibinfo {pages} {2117} (\bibinfo {year}
  {1994})}\BibitemShut {NoStop}%
\bibitem [{\citenamefont {Paul}\ \emph {et~al.}(2001)\citenamefont {Paul},
  \citenamefont {Toma}, \citenamefont {Breger}, \citenamefont {Mullot},
  \citenamefont {Aug{\'e}}, \citenamefont {Balcou}, \citenamefont {Muller},\
  and\ \citenamefont {Agostini}}]{paul2001observation}%
  \BibitemOpen
  \bibfield  {author} {\bibinfo {author} {\bibfnamefont {P.~.~M.}\ \bibnamefont
  {Paul}}, \bibinfo {author} {\bibfnamefont {E.}~\bibnamefont {Toma}}, \bibinfo
  {author} {\bibfnamefont {P.}~\bibnamefont {Breger}}, \bibinfo {author}
  {\bibfnamefont {G.}~\bibnamefont {Mullot}}, \bibinfo {author} {\bibfnamefont
  {F.}~\bibnamefont {Aug{\'e}}}, \bibinfo {author} {\bibfnamefont
  {P.}~\bibnamefont {Balcou}}, \bibinfo {author} {\bibfnamefont
  {H.}~\bibnamefont {Muller}}, \ and\ \bibinfo {author} {\bibfnamefont
  {P.}~\bibnamefont {Agostini}},\ }\href@noop {} {\bibfield  {journal}
  {\bibinfo  {journal} {Science}\ }\textbf {\bibinfo {volume} {292}},\ \bibinfo
  {pages} {1689} (\bibinfo {year} {2001})}\BibitemShut {NoStop}%
\bibitem [{\citenamefont {Hentschel}\ \emph {et~al.}(2001)\citenamefont
  {Hentschel}, \citenamefont {Kienberger}, \citenamefont {Spielmann},
  \citenamefont {Reider}, \citenamefont {Milosevic}, \citenamefont {Brabec},
  \citenamefont {Corkum}, \citenamefont {Heinzmann}, \citenamefont {Drescher},\
  and\ \citenamefont {Krausz}}]{hentschel2001attosecond}%
  \BibitemOpen
  \bibfield  {author} {\bibinfo {author} {\bibfnamefont {M.}~\bibnamefont
  {Hentschel}}, \bibinfo {author} {\bibfnamefont {R.}~\bibnamefont
  {Kienberger}}, \bibinfo {author} {\bibfnamefont {C.}~\bibnamefont
  {Spielmann}}, \bibinfo {author} {\bibfnamefont {G.~A.}\ \bibnamefont
  {Reider}}, \bibinfo {author} {\bibfnamefont {N.}~\bibnamefont {Milosevic}},
  \bibinfo {author} {\bibfnamefont {T.}~\bibnamefont {Brabec}}, \bibinfo
  {author} {\bibfnamefont {P.}~\bibnamefont {Corkum}}, \bibinfo {author}
  {\bibfnamefont {U.}~\bibnamefont {Heinzmann}}, \bibinfo {author}
  {\bibfnamefont {M.}~\bibnamefont {Drescher}}, \ and\ \bibinfo {author}
  {\bibfnamefont {F.}~\bibnamefont {Krausz}},\ }\href@noop {} {\bibfield
  {journal} {\bibinfo  {journal} {Nature}\ }\textbf {\bibinfo {volume} {414}},\
  \bibinfo {pages} {509} (\bibinfo {year} {2001})}\BibitemShut {NoStop}%
\bibitem [{\citenamefont {Krausz}\ and\ \citenamefont
  {Ivanov}(2009)}]{krausz2009attosecond}%
  \BibitemOpen
  \bibfield  {author} {\bibinfo {author} {\bibfnamefont {F.}~\bibnamefont
  {Krausz}}\ and\ \bibinfo {author} {\bibfnamefont {M.}~\bibnamefont
  {Ivanov}},\ }\href@noop {} {\bibfield  {journal} {\bibinfo  {journal}
  {Reviews of Modern Physics}\ }\textbf {\bibinfo {volume} {81}},\ \bibinfo
  {pages} {163} (\bibinfo {year} {2009})}\BibitemShut {NoStop}%
\bibitem [{\citenamefont {W{\"o}rner}\ \emph {et~al.}(2010)\citenamefont
  {W{\"o}rner}, \citenamefont {Bertrand}, \citenamefont {Kartashov},
  \citenamefont {Corkum},\ and\ \citenamefont
  {Villeneuve}}]{worner2010following}%
  \BibitemOpen
  \bibfield  {author} {\bibinfo {author} {\bibfnamefont {H.~J.}\ \bibnamefont
  {W{\"o}rner}}, \bibinfo {author} {\bibfnamefont {J.~B.}\ \bibnamefont
  {Bertrand}}, \bibinfo {author} {\bibfnamefont {D.~V.}\ \bibnamefont
  {Kartashov}}, \bibinfo {author} {\bibfnamefont {P.~B.}\ \bibnamefont
  {Corkum}}, \ and\ \bibinfo {author} {\bibfnamefont {D.~M.}\ \bibnamefont
  {Villeneuve}},\ }\href@noop {} {\bibfield  {journal} {\bibinfo  {journal}
  {Nature}\ }\textbf {\bibinfo {volume} {466}},\ \bibinfo {pages} {604}
  (\bibinfo {year} {2010})}\BibitemShut {NoStop}%
\bibitem [{\citenamefont {Lutman}\ \emph {et~al.}(2016)\citenamefont {Lutman},
  \citenamefont {Maxwell}, \citenamefont {MacArthur}, \citenamefont {Guetg},
  \citenamefont {Berrah}, \citenamefont {Coffee}, \citenamefont {Ding},
  \citenamefont {Huang}, \citenamefont {Marinelli}, \citenamefont {Moeller}
  \emph {et~al.}}]{lutman2016fresh}%
  \BibitemOpen
  \bibfield  {author} {\bibinfo {author} {\bibfnamefont {A.~A.}\ \bibnamefont
  {Lutman}}, \bibinfo {author} {\bibfnamefont {T.~J.}\ \bibnamefont {Maxwell}},
  \bibinfo {author} {\bibfnamefont {J.~P.}\ \bibnamefont {MacArthur}}, \bibinfo
  {author} {\bibfnamefont {M.~W.}\ \bibnamefont {Guetg}}, \bibinfo {author}
  {\bibfnamefont {N.}~\bibnamefont {Berrah}}, \bibinfo {author} {\bibfnamefont
  {R.~N.}\ \bibnamefont {Coffee}}, \bibinfo {author} {\bibfnamefont
  {Y.}~\bibnamefont {Ding}}, \bibinfo {author} {\bibfnamefont {Z.}~\bibnamefont
  {Huang}}, \bibinfo {author} {\bibfnamefont {A.}~\bibnamefont {Marinelli}},
  \bibinfo {author} {\bibfnamefont {S.}~\bibnamefont {Moeller}},  \emph
  {et~al.},\ }\href@noop {} {\bibfield  {journal} {\bibinfo  {journal} {Nature
  Photonics}\ }\textbf {\bibinfo {volume} {10}},\ \bibinfo {pages} {745}
  (\bibinfo {year} {2016})}\BibitemShut {NoStop}%
\bibitem [{\citenamefont {Zholents}(2005)}]{zholents2005method}%
  \BibitemOpen
  \bibfield  {author} {\bibinfo {author} {\bibfnamefont {A.~A.}\ \bibnamefont
  {Zholents}},\ }\href@noop {} {\bibfield  {journal} {\bibinfo  {journal}
  {Physical Review Special Topics-Accelerators and Beams}\ }\textbf {\bibinfo
  {volume} {8}},\ \bibinfo {pages} {040701} (\bibinfo {year}
  {2005})}\BibitemShut {NoStop}%
\bibitem [{\citenamefont {Hemsing}\ \emph {et~al.}(2014)\citenamefont
  {Hemsing}, \citenamefont {Stupakov}, \citenamefont {Xiang},\ and\
  \citenamefont {Zholents}}]{hemsing2014beam}%
  \BibitemOpen
  \bibfield  {author} {\bibinfo {author} {\bibfnamefont {E.}~\bibnamefont
  {Hemsing}}, \bibinfo {author} {\bibfnamefont {G.}~\bibnamefont {Stupakov}},
  \bibinfo {author} {\bibfnamefont {D.}~\bibnamefont {Xiang}}, \ and\ \bibinfo
  {author} {\bibfnamefont {A.}~\bibnamefont {Zholents}},\ }\href@noop {}
  {\bibfield  {journal} {\bibinfo  {journal} {Reviews of Modern Physics}\
  }\textbf {\bibinfo {volume} {86}},\ \bibinfo {pages} {897} (\bibinfo {year}
  {2014})}\BibitemShut {NoStop}%
\bibitem [{\citenamefont {Prat}\ \emph {et~al.}(2015)\citenamefont {Prat},
  \citenamefont {L{\"o}hl},\ and\ \citenamefont {Reiche}}]{prat2015efficient}%
  \BibitemOpen
  \bibfield  {author} {\bibinfo {author} {\bibfnamefont {E.}~\bibnamefont
  {Prat}}, \bibinfo {author} {\bibfnamefont {F.}~\bibnamefont {L{\"o}hl}}, \
  and\ \bibinfo {author} {\bibfnamefont {S.}~\bibnamefont {Reiche}},\
  }\href@noop {} {\bibfield  {journal} {\bibinfo  {journal} {Physical Review
  Special Topics-Accelerators and Beams}\ }\textbf {\bibinfo {volume} {18}},\
  \bibinfo {pages} {100701} (\bibinfo {year} {2015})}\BibitemShut {NoStop}%
\bibitem [{\citenamefont {Huang}\ \emph {et~al.}(2017)\citenamefont {Huang},
  \citenamefont {Ding}, \citenamefont {Feng}, \citenamefont {Hemsing},
  \citenamefont {Huang}, \citenamefont {Krzywinski}, \citenamefont {Lutman},
  \citenamefont {Marinelli}, \citenamefont {Maxwell},\ and\ \citenamefont
  {Zhu}}]{huang2017generating}%
  \BibitemOpen
  \bibfield  {author} {\bibinfo {author} {\bibfnamefont {S.}~\bibnamefont
  {Huang}}, \bibinfo {author} {\bibfnamefont {Y.}~\bibnamefont {Ding}},
  \bibinfo {author} {\bibfnamefont {Y.}~\bibnamefont {Feng}}, \bibinfo {author}
  {\bibfnamefont {E.}~\bibnamefont {Hemsing}}, \bibinfo {author} {\bibfnamefont
  {Z.}~\bibnamefont {Huang}}, \bibinfo {author} {\bibfnamefont
  {J.}~\bibnamefont {Krzywinski}}, \bibinfo {author} {\bibfnamefont
  {A.}~\bibnamefont {Lutman}}, \bibinfo {author} {\bibfnamefont
  {A.}~\bibnamefont {Marinelli}}, \bibinfo {author} {\bibfnamefont
  {T.}~\bibnamefont {Maxwell}}, \ and\ \bibinfo {author} {\bibfnamefont
  {D.}~\bibnamefont {Zhu}},\ }\href@noop {} {\bibfield  {journal} {\bibinfo
  {journal} {Physical review letters}\ }\textbf {\bibinfo {volume} {119}},\
  \bibinfo {pages} {154801} (\bibinfo {year} {2017})}\BibitemShut {NoStop}%
\bibitem [{\citenamefont {Behrens}\ \emph {et~al.}(2014)\citenamefont {Behrens}
  \emph {et~al.}}]{behrens14etal}%
  \BibitemOpen
  \bibfield  {author} {\bibinfo {author} {\bibfnamefont {C.}~\bibnamefont
  {Behrens}} \emph {et~al.},\ }\href@noop {} {\bibfield  {journal} {\bibinfo
  {journal} {Nature Communications}\ }\textbf {\bibinfo {volume} {5}},\
  \bibinfo {pages} {3762} (\bibinfo {year} {2014})}\BibitemShut {NoStop}%
\bibitem [{\citenamefont {Hoegner}(2015)}]{HHGmaxcode}%
  \BibitemOpen
  \bibfield  {author} {\bibinfo {author} {\bibfnamefont {M.}~\bibnamefont
  {Hoegner}},\ }\href@noop {} {\enquote {\bibinfo {title} {Hhgmax},}\ }\bibinfo
  {howpublished} {\url{http://github.com/Leberwurscht/HHGmax}} (\bibinfo {year}
  {2015})\BibitemShut {NoStop}%
\bibitem [{\citenamefont {Doumy}\ \emph {et~al.}(2009)\citenamefont {Doumy},
  \citenamefont {Wheeler}, \citenamefont {Roedig}, \citenamefont {Chirla},
  \citenamefont {Agostini},\ and\ \citenamefont
  {DiMauro}}]{doumy2009attosecond}%
  \BibitemOpen
  \bibfield  {author} {\bibinfo {author} {\bibfnamefont {G.}~\bibnamefont
  {Doumy}}, \bibinfo {author} {\bibfnamefont {J.}~\bibnamefont {Wheeler}},
  \bibinfo {author} {\bibfnamefont {C.}~\bibnamefont {Roedig}}, \bibinfo
  {author} {\bibfnamefont {R.}~\bibnamefont {Chirla}}, \bibinfo {author}
  {\bibfnamefont {P.}~\bibnamefont {Agostini}}, \ and\ \bibinfo {author}
  {\bibfnamefont {L.}~\bibnamefont {DiMauro}},\ }\href@noop {} {\bibfield
  {journal} {\bibinfo  {journal} {Physical review letters}\ }\textbf {\bibinfo
  {volume} {102}},\ \bibinfo {pages} {093002} (\bibinfo {year}
  {2009})}\BibitemShut {NoStop}%
\end{thebibliography}%

\end{document}